\documentclass[aps]{revtex4}     %
\usepackage{amsmath,amsthm,latexsym,amssymb,bm,mathbbol,graphicx,subfigure}
\renewcommand{\v}[1]{\mathbf{#1}}
\renewcommand{\v}[1]{\mathbf{#1}}\newcommand{\na}[0]{\nabla}

\newcommand{\wh}[1]{\widehat{#1}}
\newcommand{\C}[1]{\mathcal{#1}}

\newcommand{\RR}[0]{\mathbb{R}}

\DeclareMathOperator{\tr}{tr}
\newenvironment{mat}[1]{\begin{array}{@{}*{#1}{r@{}l}@{}}}{\end{array}}
\begin{document}
\title{Reconstruction of spatially inhomogeneous dielectric tensors
via optical tomography}
\author{Hanno Hammer\email{H.Hammer@umist.ac.uk}}
\author{William R. B. Lionheart\email{Bill.Lionheart@umist.ac.uk}}
\affiliation{Department of Mathematics, \\ University of Manchester
Institute of Science and Technology (UMIST), \\ PO Box 88, Manchester
M60 1QD, United Kingdom}
\begin{abstract}
A method to reconstruct weakly anisotropic inhomogeneous dielectric
tensors inside a transparent medium is proposed. The mathematical
theory of Integral Geometry is cast into a workable framework which
allows the full determination of dielectric tensor fields by scalar
Radon inversions of the polarization transformation data obtained from
six planar tomographic scanning cycles. Furthermore, a careful
derivation of the usual equations of integrated photoelasticity in
terms of heuristic length scales of the material inhomogeneity and
anisotropy is provided, making the paper a self-contained account
about the reconstruction of arbitrary three-dimensional, weakly
anisotropic dielectric tensor fields.

\vspace{1em}
OCIS numbers 100.3190, 160.1190, 080.2710
\end{abstract}

\maketitle 

\section{Introduction}

The inverse boundary value problem of recovery of anisotropic
spatially varying electromagnetic properties of a medium from external
measurements at a fixed frequency is among the most mathematically
challenging of inverse problems. For low-frequency electromagnetic
measurements where a static approximation is valid, an anisotropic
dielectric tensor or conductivity tensor is uniquely determined by
complete surface electrical measurements up to a gauge
condition\cite{LassasUhlmann}. For isotropic\cite{OPS} and chiral
isotropic media\cite{JoshiMcD} a knowledge of all pairs of tangential
electric and magnetic fields at the boundary, for a single
non-exceptional frequency, is known to be sufficient to recover the
material properties. Anisotropic materials are important in many
practical problems with anisotropy arising from, for example, flow,
deformation, crystal or liquid crystal structure, and effective
anisotropic properties arising from homogenization of fibrous or
layered composite materials. For general anisotropic media the
question of sufficiency of data for reconstruction remains open.

In this paper we concentrate on a specific high-frequency case of
considerable practical importance. We assume that the material is
non-magnetic, i.e. has a homogeneous and isotropic permeability equal
to the vacuum value $\mu_0$; that the conductivity is negligible; and
that the permittivity, or dielectric tensor, is {\em weakly
anisotropic} in a sense that will be defined below.  We present a
mathematical framework which allows us to invert the polarization
transformation data obtained from tomographic measurements of light
rays passing through an optically anisotropic material at sufficiently
many angles for the six independent components of the (spatially
varying) dielectric tensor $\epsilon_{ij}$ inside the specimen. It
will be shown that, for weak anisotropy, our method allows the full
determination of all tensor components, provided that tomographic
measurements are made for six carefully chosen orientations of the
planes in which the light rays scan the medium.

The equations describing the passage of light
through inhomogeneous and weakly anisotropic media have been
formulated in Refs.~\onlinecite{Kravtsov1968a,KravtsovOrlov1980,FKN},
and in section~\ref{Scales} we shall show how to obtain these
equations from a geometric-optical starting point by expanding the
electric field in powers of appropriate length scales which describe
the inhomogeneity and anisotropy of the medium. Also, a condition for
``weak'' anisotropy will be specified 
upon which we shall
linearize (section~\ref{Linear}) the equations determining the
polarization transfer matrix along the light rays. In
section~\ref{6Radon} we then present our method of reconstructing the
permittivity tensor in the linearized limit by performing scalar Radon
transforms on the polarization transformation data obtained from six
scanning cycles on different planes intersecting the specimen. In
section~\ref{Example} the method is demonstrated by giving a visual
example of reconstruction of the permittivity tensor inside an axially
loaded cylindrical bar: in this case, the stress tensor in the
(transparent) medium gives rise to optical anisotropy, and it will be
seen that our method works well in the limit of weak anisotropy. The
mathematical basis for this method has been anticipated in the seminal
work of Sharafutdinov\cite{Sharafutdinov}, whose book contains a
general theory of ``ray transforms'' of tensor fields on
$n$-dimensional Euclidean spaces, and an examination into the
possibility to invert them for reconstructing the underlying tensor
fields. Our work presented here is an adaption, and partial
reformulation, of this highly mathematical framework, to the specific
objective of reconstructing inhomogeneous dielectric tensor fields via
optical tomography.

It should be mentioned that the determination of anisotropic
dielectric tensors as presented here has a closely related variant in
the problem of reconstruction of a stress tensor field inside a loaded
transparent material; the phenomenon that an initially isotropic
medium becomes optically anisotropic when under strain is called {\it
Photoelasticity} \cite{Frocht, CokerFilon, TheocarisEA, JessopHarris,
Hendry} and may be used to obtain information about the internal
stress in a strained medium from polarization transformation data
obtained by tomographical methods. The photoelastic effect as a means
to retrieve information about internal stresses has been studied
extensively: A method termed ``Integrated
Photoelasticity''\cite{Aben1966a, Aben1979, AbenEA1989a,
AbenGuillemet, AbenEA1997a} is well-known, and it was pointed
out\cite{Davin1969a, AbenIdnurmPuro1990a, AbenJosepsonKellPuro1991a,
AbenErrapartAinolaAnton2003a} that information about the {\it
difference} of principal stress components could be retrieved from
appropriate Radon transformations of polarization transformation
data. However, these methods do not succeed in reconstructing the
stress components separately and therefore the full stress tensor, and
the linearly related dielectric tensor, can be obtained in this way
only for systems exhibiting a certain degree of symmetry, such as an
axial symmetry. Other methods of reconstruction have been suggested,
for example, a three-beam measurement scheme\cite{AndrienkoEA1994a},
where for axisymmetric systems an onion-peeling reconstruction
algorithm was proposed\cite{AndrienkoEA1992a,
AndrienkoEA1992b}. Another method, which in principle is capable of
determining a general three-dimensional permittivity tensor, is the
``load incremental approach''\cite{WijerathneEA2002a}. Here, the
stress on the object is increased in small increments, and at each
step, a measurement cycle is performed.

The new results in this paper are twofold: Firstly, we derive the set
of equations determining the polarization transfer matrices for a
given light ray scanning the object. These equations are somewhat
related to the standard equations of integrated
photoelasticity\cite{Aben1979}, but here we present a more rigorous
exposition of the heuristic length scales\cite{FKN}, and the
approximations based thereupon, which underlie the usual derivation of
these equations from Maxwell's equations in a material medium; this is
done in sections~\ref{Scales}, \ref{TransferMatrix}
and~\ref{Linear}. Secondly, we present a novel scheme for
reconstruction of arbitrarily inhomogeneous dielectric tensors in the
interior of the specimen, subject only to the condition that the
birefringence is ``weak'' in a sense which will be specified in
sections~\ref{6Radon} and~\ref{Example}.

\section{Heuristic length scales} \label{Scales}

It is assumed that the material is non-absorbing for optical
wavelengths, and furthermore is non-magnetic, $\mu_{ij} = \mu_0
\delta_{ij}$, where $\mu_0$ is the magnetic permeability of vacuum. As
for the permittivity we assume that the dielectric tensor deviates
``weakly'' from a global spatial average
\begin{equation}
\label{average1}
 \epsilon = \frac{1}{ {\rm vol}\, B} \int_B d^3\!x\; \frac{1}{3} \tr(
 \underline{\epsilon}) \quad,
\end{equation}
where $B$ denotes the body having a volume ${\rm vol}~\!B$, $\tr
\underline{\epsilon} = \sum_{i=1}^3 \epsilon_{ii}$ is the trace
operation, and $\underline{ \epsilon} = \underline{ \epsilon}(\v{x})$
is the dielectric tensor. In a zero-order approximation the material
therefore may be regarded as homogeneously isotropic, with scalar
dielectric constant $\epsilon$ as defined
in~(\ref{average1}). Typically, this behaviour of weak deviation from
a homogeneously isotropic background will be satisfied for glasses and
certain resins under moderate internal stress or external load. The
scalar constant of permittivity $\epsilon$ will be a reference
quantity when we specify the dimensionless degree of anisotropy in
eq.~(\ref{scale5}).

For the actual problems relevant to our work, the length scales
characterising inhomogeneities in the material are much larger than
the wavelength of the (monochromatic) light passing through the
object, so that the usage of geometric-optical approximations is
justified. Heuristically, two such length scales can be
conceptualized\cite{FKN}: A scale $l_0$ characterising the degree of
inhomogeneity may be introduced by
\begin{equation}
\label{scales1}
 l_0\, | \bm{\kappa} \cdot \na \tr \underline{\epsilon}(\v{x})| \sim |
 \tr \underline{\epsilon}(\v{x})| \quad,
\end{equation}
where $\bm{\kappa}$ denotes a unit vector in the direction of wave
propagation. Furthermore, in any anisotropic medium, two preferred
polarization directions $\v{e}_p(\v{x})$, $p=1,2$, for each given
direction of wave propagation\cite{BornWolf, Fowles, SommerfeldOptics,
Ditchburn, Longhurst}, exist at each point, so that a scale $l_p$
measuring the rate of variation in these polarization directions can
be speficied by
\begin{equation}
\label{scale2}
 l_p\, | \bm{\kappa} \cdot \na \v{e}_p(\v{x})| \sim | \v{e}_p(\v{x})|
 \quad.
\end{equation}
These scales should be compared to the average wavelength
$\overline{\lambda}$ of the monochromatic light passing through the
medium, so that, when the limit
\begin{equation}
\label{GOLimit}
 l_p, \, l_0 \gg \overline{\lambda}
\end{equation}
is satisfied, the (complex) electric field may be given in the form
\begin{equation}
\label{geomopt1}
 \wh{\v{E}}(\v{x},t) = \v{E}(\v{x})\, e^{i\phi(\v{x}) - i\omega t}
 \quad, \quad
\end{equation}
where the eikonal $\phi(\v{x}+ \overline{\lambda} \v{s}) = \phi(\v{x})
+ \overline{\lambda} \v{s} \cdot \na \phi(\v{x}) + \C{O}\left(
\overline{\lambda} / l_0 \right)$ describes a locally-plane wave with
wave vector $\na \phi(\v{x})$, and the amplitude $\overline{\lambda}\,
\bm{\kappa} \cdot \na \v{E} \sim \C{O} \left( \overline{\lambda} / l_p
\right)\, \v{E}$ varies weakly on the length scale
$\overline{\lambda}$. Motivated by these considerations, Fuki,
Kravtsov, Naida (FKN)\cite{FKN} introduce a dimensionless scale
%
\begin{equation}
\label{scale3}
 \alpha = \max \left\{ \frac{\overline{\lambda}}{l_0},\,
 \frac{\overline{\lambda}}{l_p} \right\} \quad.
\end{equation}
The limit of geometrical optics then can be specified by the condition
\begin{equation}
\label{geomopt2}
 \alpha \ll 1 \quad.
\end{equation}

We also need an explicit measure of anisotropy
\begin{subequations}
\label{scale5}
\begin{align}
 A_{ij} & = \frac{ \epsilon_{ij} - \delta_{ij} \epsilon}{\epsilon}
 \quad, \label{scale5a} \\
 \beta & \sim \max \left\| A_{ij} \right\| \quad, \label{scale5b}
\end{align}
\end{subequations}
where $\beta$ is an appropriate number characterising the magnitude of
the components of the dimensionless anisotropy tensor $A_{ij}$, such
as a global maximum, etc. If anisotropy is not weak, then at each
point in the medium we have a continuous splitting of rays due to the
fact that there are two distinct phase velocities, and two distinct
ray velocities, for any given propagation direction. A condition for
weak anisotropy therefore arises if we demand that the passage of
light through the material can be described by a {\it single} ray
which is influenced by the local variations of the optical tensors
only in the way that the polarization directions rotate. This is the
situation that commonly occurs in photoelasticity and is also of
greatest interest to our work.

It was shown by FKN\cite{FKN} that ray splitting can be ignored, if
\begin{equation}
\label{elec5}
 \frac{\beta}{\alpha } \lesssim 1 \quad.
\end{equation}
In this case\cite{FKN} ``it is impossible to discriminate
experimentally between split rays'', and one can effectively replace
the two rays by a single {\it isotropic} ray which is obtained from
the isotropic part of the dielectric tensor alone. This is indeed the
domain we are most concerned with, since experimentally no ray
splitting is seen in photoelastic experiments. In fact, for those
applications we are interested in it is usually true that light
propagates along straight lines through the specimen, so that the
trial solution~(\ref{geomopt1}) may be replaced by an even stronger
ansatz
\begin{equation}
\label{elec8}
 \bm{\C{E}}(\v{x},t) = \v{E}(\v{x})\, e^{i \v{k} \cdot \v{x} - i\omega
 t} \quad
\end{equation}
with {\it constant} wave vector $\v{k}$, just as for a plane wave. The
phase velocity associated with this wave vector is the one associated
with the average permittivity $\epsilon$ defined in
eq.~(\ref{average1}),
\begin{equation}
\label{elec9}
 k = \frac{\omega}{u} \quad, \quad u = \frac{1}{\sqrt{\mu_0 \epsilon}}
\quad, \quad \lambda = \frac{2\pi}{k} \quad.
\end{equation}
However, the amplitude $\v{E}(\v{x})$ has a spatial dependence which
accounts for the variation of the two preferred polarization
directions along the light ray.

Under the conditions~(\ref{geomopt2}) and (\ref{elec5}), the electric
field $\v{E}$ and electric displacement $\v{D}$ behave like
\begin{equation}
\label{elec10}
 \bm{\kappa} \cdot \v{D} \sim \C{O}\left( \frac{\lambda}{l_p} \right)\,
 \v{D} \quad, \quad \bm{\kappa} \cdot \v{E} \sim \C{O}\left(
 \frac{\lambda}{l_p} + \beta \right)\, \v{E} \quad.
\end{equation}
This means that $\v{D}$ is nearly transverse, while the same is true
for $\v{E}$ only if we assume in addition that the degree of
anisotropy is small,
\begin{equation}
\label{elec11}
 \beta \ll 1 \quad.
\end{equation}
This is the condition of {\it weak anisotropy}, and our method is
formulated for this ``quasi-isotropic'' regime.

\section{Equations satisfied by the transfer matrices} \label{TransferMatrix}

The information about the change in the state of polarization of a
light beam passing through the material is encoded in a
two-dimensional unitary {\it transfer matrix}. The equation satisfied
by the transfer matrix along a light ray is given in most accounts on
photoelastic tomography\cite{Davin1969a, AbenIdnurmPuro1990a,
AbenJosepsonKellPuro1991a,
AbenErrapartAinolaAnton2003a,AndrienkoEA1994a,AndrienkoEA1992a,
AndrienkoEA1992b}, but the various approximations taken in the process
of neglecting higher powers of ratios $\lambda/l_0$ and $\lambda/l_p$
are not always stated clearly; we therefore briefly summarize the
necessary steps here:

On inserting (\ref{elec8}) into Maxwell's equations we obtain
\begin{equation}
\label{dyneq1}
 \v{k} \times \bigg[ \v{k} \times \v{E} \bigg] + \omega^2 \mu_0
 \underline{\epsilon} \v{E} - i \bigg[ \na \times \left( \v{k} \times
 \v{E} \right) + \v{k} \times \left( \na \times \v{E} \right) \bigg] -
 \na \times \big( \na \times \v{E} \big) = 0 \quad.
\end{equation}
It is easy to show that
\begin{equation}
\label{elec12}
 \na \times \na \times \v{E} \sim \C{O}\left(\rule{0pt}{18pt} \left(
 \frac{\lambda}{l_p} \right)^2 \right)\, \v{E}
\end{equation}
hence the term can be neglected in the geometrical-optical
limit~(\ref{geomopt2}). Then~(\ref{dyneq1}) takes the form
\begin{equation}
\label{poltransfer1}
 \bm{\kappa} \times \Big( \bm{\kappa} \times \v{E} \Big) - \frac{i}{k}
 \Big\{ \na \times \big( \bm{\kappa} \times \v{E} \big) + \bm{\kappa}
 \times \big( \na \times \v{E} \big) \Big\} + \mu_0 u^2
 \underline{\epsilon} \v{E} = 0 \quad,
\end{equation}
where $u$ was given in eq.~(\ref{elec9}). The longitudinal component
of eq.~(\ref{poltransfer1}), obtained by projection onto the unit
vector $\bm{\kappa} = \v{k}/k$ in the direction of propagation of the
light beam, is of the order
\begin{equation}
\label{elec15}
 \C{O}\left( \frac{\lambda}{l_p} \right) \v{E}
\end{equation}
and is neglected in the geometrical-optical limit. Hence we only
retain the transverse components of $\v{E}$ and $\v{D}$, i.e. the
components perpendicular to the wave propagation $\bm{\kappa}$. 

Now let us study eq.~(\ref{poltransfer1}) in a coordinate system in
which the direction of propagation $\bm{\kappa}$ is along the $z$
axis. Then~(\ref{poltransfer1}) becomes
\begin{equation}
\label{elec16}
\frac{d}{dz} \left[ \begin{mat}{1} & E_1 \\ & E_2 \end{mat} \right] =
i \frac{\pi}{\lambda} \left[ \begin{mat}{2} & A_{11} && A_{12} \\ &
A_{21} && A_{22} \end{mat} \right] \left[ \begin{mat}{1} & E_1 \\ &
E_2 \end{mat} \right] \quad,
\end{equation}
where $A_{ij}$ was defined in eq.~(\ref{scale5a}). The solution
of~(\ref{elec16}) can be expressed via a transfer matrix
\begin{equation}
\label{elec17}
 \left[ \begin{mat}{1} & E_1(z) \\ & E_2(z) \end{mat} \right] =
 U(z,z_0) \left[ \begin{mat}{1} & E_1(z_0) \\ & E_2(z_0) \end{mat}
 \right] \quad,
\end{equation}
where $U$ satisfies an ordinary differential equation similar to
(\ref{elec16}), together with initial condition
\begin{equation}
\label{elec18}
\begin{aligned}
 \frac{d}{dz} U(z,z_0) & = i \frac{\pi}{\lambda}\, A_{\bot}(z)\,
 U(z,z_0) \quad, \\[10pt]
 U(z_0,z_0) & = \Eins_2 = \left( \begin{mat}{2} & 1 && 0 \\ & 0 && 1
 \end{mat} \right)  \quad.
\end{aligned}
\end{equation}
Eqs.~(\ref{elec18}) can be expressed as an integral equation
\begin{equation}
\label{elec19}
 U(z,z_0) = \Eins_2 + i \frac{\pi}{\lambda} \int\limits_{z_0}^{z}
 dz_1\; A_{\bot}(z_1)\, U(z_1, z_0) \quad,
\end{equation}
where $A_{\bot}$ denotes the matrix of transverse components of
$A_{ij}$ as they appear in eq.~(\ref{elec16}). A formal solution
of~(\ref{elec19}) is given by the Born-Neumann series
\begin{equation}
\label{elec20}
 U(z,z_0) = \Eins_2 + \left( i \frac{\pi}{\lambda} \right)
 \int\limits_{z_0}^z dz_1\; A_{\bot}(z_1) + \left( i
 \frac{\pi}{\lambda} \right)^2 \int\limits_{z_0}^z dz_1\;
 A_{\bot}(z_1) \int\limits_{z_0}^{z_1} dz_2\; A_{\bot}(z_2) + \cdots
\end{equation}

The transfer matrix $U$ is unitary and thus preserves the norm of the
complex electric field vector. Physically this means that intensity is
preserved, so unitarity here just expresses energy conservation of the
light ray. This must indeed be the case, since we have assumed a
non-absorbing medium.

\section{The linearized inverse problem} \label{Linear}

Assuming that the transfer matrices $U$ have been determined for
sufficiently many light rays scanning the medium, the associated
inverse problem now consists in reconstructing the anisotropy tensor
$A_{ij}$ from the collection of these transfer matrices; this inverse
problem is obviously non-linear in $A_{ij}$, as can be seen from
eqs.~(\ref{elec19} and \ref{elec20}). The solution to the fully
non-linear problem is not known as yet. However, in the
quasi-isotropic regime we can deal with the linearized inverse
problem: this is defined by a truncation of the Born-Neumann series in
(\ref{elec20}) after the first-order term
\begin{equation}
\label{linearized1}
 U(z,z_0) = \Eins_2 + i \frac{\pi}{\lambda} \int\limits_{z_0}^z dz_1\;
 A_{\bot}(z_1) \quad.
\end{equation}
For example, for a relative anisotropy of $\nu \sim 10^{-9}$, a
wavelength of $\lambda \sim 0.5 \times 10^{-6}$ {\rm m} and assuming a
length of $L\sim 1$~{\rm m} of the object, we find that the
first-order term in (\ref{linearized1}) is of the order $10^{-3}$,
hence the linearization will be a good approximation in this case.

The transfer matrices $U$ must be determined by suitable measurements
of the change of polarization along each light ray passing through the
medium at many different angles, possibly including interferometric
methods. By measuring three so-called characteristic
parameters\cite{Aben1966a} the $SU(2)$-factor of the transfer matrices
$U$ can be computed using the Poincar\'e equivalence
theorem\cite{Poincare1892}, a matrix decomposition theorem which
allows to interprete the characteristic parameters as the retardation
angle $\Delta$, the angle of the fast axis $\theta$, and the rotation
angle $\delta$ of an equivalent optical model consisting of a linear
retarder and a rotator with these optical parameters. The Poincar\'e
equivalence theorem can be formulated in terms of Jones matrices or
Stokes parameters on the Poincar\'e sphere; a contemporary exposition
of these relations was given recently in
Ref.~\onlinecite{HammerJModOpt2004a}.

Thus, the measurement of $\Delta, \theta, \delta$ determines a
unimodular matrix $ S(\Delta, \theta, \delta)$ such that
\begin{equation}
\label{unimo1}
 U = S(\Delta, \theta, \delta) \times e^{i \Phi} \quad, \quad
 S(\Delta, \theta, \delta) \in SU(2) \quad,
\end{equation}
where $\Phi$ is the {\it global phase} of the transfer matrix $U$. In
the general case, assuming no restrictions on the degree of
anisotropy, this global phase can be arbitrarily large, and
furthermore can{\bf not} be determined from the characteristic
parameters alone, but must be measured e.g. through interferometric
methods, for each given light ray. However, in the limit of weak
anisotropy, the global phase is effectively determined by the
characteristic parameters alone: Suppose that the unimodular matrix
$S$ has been computed by measuring the parameters $\Delta, \theta,
\delta$ using the Poincar\'e equivalence theorem; from
eq.~(\ref{linearized1}) it then follows that the unknown phase $\Phi$
must be chosen so as to make the real parts of the diagonal
(non-diagonal) matrix elements equal to one (zero),
\begin{equation}
\label{unimo2}
\begin{aligned}
 \Re \bigg( e^{i \Phi}\, S_{11} \bigg) & = \Re \bigg( e^{i \Phi}\,
 S_{22} \bigg) = 1 \quad, \\
 \Re \bigg( e^{i \Phi}\, S_{12} \bigg) & = \Re \bigg( e^{i \Phi}\,
 S_{21} \bigg) = 0 \quad.
\end{aligned}
\end{equation}
We can therefore determine the phase $\Phi$ from any of these
equations, or, for the sake of numerical stability, from all of them,
so as to obtain an average value of $\Phi$. Thus, in the
weak-anisotropy limit, the number of real ``degrees of freedom'' in
the transfer matrices is effectively equal to three, rather than four
as in the general case.

\section{Solution of the linearized inverse problem by six scalar
Radon inversions} \label{6Radon}

We now show that the linearized inverse problem can be reduced to six
scalar Radon inversions performed on the polarization transformation
data. We first specify a plane $P(\v{y},\bm{\eta})$ in $\RR^3$ which
intersects the specimen and contains the point $\v{y}$; the
orientation of the plane is determined by a unit vector $\bm{\eta}$
normal to the plane. Consider the straight line $t \mapsto \v{y} + t
\bm{\kappa}$ with unit vector $\bm{\kappa}$ lying in
$P(\v{y},\bm{\eta})$, describing a light ray passing through the
specimen and lying in the given plane $P(\v{y},\bm{\eta})$. Let
$\bm{\xi}$ be a third unit vector, orthogonal to $\bm{\kappa}$ and
$\bm{\eta}$ in such a way that $(\bm{\xi}, \bm{\eta}, \bm{\kappa})$
form a right-handed system. Then the equation describing the
polarization transformation along the light ray in the direction
$\bm{\kappa}$ is given by the analogue of~(\ref{linearized1}),
\begin{equation}
\label{linearized2}
 \left[ \begin{mat}{2} & U_{\xi\xi}(t,t_0) && U_{\xi\eta}(t,t_0) \\ &
 U_{\eta\xi}(t,t_0) && U_{\eta\eta}(t,t_0) \end{mat} \right] = \left[
 \begin{mat}{2} & 1 && 0 \\ & 0 && 1 \end{mat} \right] + i
 \frac{\pi}{\lambda} \int\limits_{t_0}^z dt_1\, \left[ \begin{mat}{2}
 & A_{\xi\xi}(t_1) && A_{\xi\eta}(t_1) \\ & A_{\eta\xi}(t_1) &&
 A_{\eta\eta}(t_1) \end{mat} \right] \quad.
\end{equation}
The measurement of characteristic parameters determines the transfer
matrix on the left-hand side of~(\ref{linearized2}). We now repeat
these measurements for all lines lying in the given plane $P(\v{y},
\bm{\eta})$ and thus obtain a collection of line integrals for the
normal component $A_{\eta\eta}$ in $P$,
\begin{equation}
\label{linearized3}
 \int dt_1\; A_{\eta\eta}(t_1) = \Big[ U_{\eta\eta}(+\infty, - \infty)
 - 1 \Big] \frac{\lambda}{i\pi}
\end{equation}
for any pair of directions $\bm{\kappa}$ and $\bm{\xi}$; we could
extend the limits of integration in~(\ref{linearized3}) to $\pm
\infty$, since in practical situations the object will be placed
inside a tank with a phase-matching fluid, hence the value of $A_{ij}$
outside the object is zero. The set of line integrals
in~(\ref{linearized3}), taken for all light rays {\it in $P$}, is
called the {\it transverse ray transform}\cite{Sharafutdinov} of
$A_{ij}$ with respect to $\bm{\eta}$.

However, the component $A_{\eta\eta} = A(\bm{\eta}, \bm{\eta})$ is
perpendicular to the plane $P$ and is therefore invariant under
$SO(2)$ rotations in that plane, so that it behaves like any other
scalar function defined on $P$. It follows that the collection of
integrals (\ref{linearized3}) is indeed the standard {\it 2D Radon
transform}\cite{HelgasonRadon} of the scalar function
$A_{\eta\eta}(\v{x})$, $\v{x} \in P(\v{y}, \bm{\eta})$, and hence can
be inverted for the component $A_{\eta\eta}$ using any numerical
scheme for Radon inversion appropriate to the circumstances, such as
filtered back-projection\cite{Natterer}. This produces the component
$A_{\eta\eta}(\v{x})$ for every point $\v{x}$ in the plane $P(\v{y},
\bm{\eta})$.

On repeating the same process for all planes parallel to
$P(\v{y},\bm{\eta})$ we reconstruct the component $A_{\eta\eta}$
within the whole specimen.

We now perform this procedure for the following six different choices
of the vector $\bm{\eta}$:
\begin{subequations}
\label{choice1}
\begin{align}
 & \bm{\eta}_1 = \v{e}_1 \quad, \quad \bm{\eta}_2 = \v{e}_2 \quad, \quad
 \bm{\eta}_3 = \v{e}_3 \quad, \label{choice1a} \\
 & \bm{\eta}_{12} = \frac{\v{e}_1 + \v{e}_2}{\sqrt{2}} \quad, \quad
 \bm{\eta}_{23} = \frac{\v{e}_3 + \v{e}_3}{\sqrt{2}} \quad, \quad
 \bm{\eta}_{31} = \frac{\v{e}_3 + \v{e}_1}{\sqrt{2}}
 \quad. \label{choice1b}
\end{align}
\end{subequations}
The scan-and-reconstruction cycle for the choices in~(\ref{choice1a})
produces components
\begin{equation}
\label{choice2}
 A_{11} \quad, \quad A_{22} \quad, \quad A_{33}
\end{equation}
of the anisotropy tensor. On the other hand, for the choices
in~(\ref{choice1b}) we obtain the following result: Let us focus
attention on the first orientation $\bm{\eta}_{12}$, where the
associated reconstruction gives us the tensor component
$A(\bm{\eta}_{12}, \bm{\eta}_{12})$ everywhere within the object. But,
due to the fact that the tensor is linear in its arguments, this
component can be expressed in the form
\begin{equation}
\label{choice3}
 A(\bm{\eta}_{12}, \bm{\eta}_{12}) = \frac{1}{2} \left( A_{11} +
 A_{22} \right) + A_{12} \quad,
\end{equation}
where we have used the fact that $A_{ij}$ is symmetric and hence
$A_{21} = A_{12}$. Thus, having already reconstructed $A_{11}$ and
$A_{22}$ everywhere in the specimen, we can immediately compute
$A_{12}$ from the reconstructed values of $A(\bm{\eta}_{12},
\bm{\eta}_{12})$ using eq.~(\ref{choice3}). On repeating this process
for the last two choices of $\bm{\eta}$ in~(\ref{choice1b}) we see
that all six tensor components of $A_{ij}$ can be reconstructed in
this way.

If the average dielectric constant $\epsilon$ of the object is known
we can compute the full dielectric tensor $\epsilon_{ij}$ immediately
using the definition~(\ref{scale5a}). On the other hand, if $\epsilon$
is not determined we can still use~(\ref{scale5a}) to write
\begin{equation}
\label{ohne1}
 \epsilon_{ij} = \epsilon \left( A_{ij} + \delta_{ij} \right) \quad,
\end{equation}
in other words, we can reconstruct $\epsilon_{ij}$ {\it up to a scale
factor} $\epsilon$; this may still produce valuable information about
the internal structure of the dielectric material, see
Fig.~\ref{figure1}.
%
%

\section{Numerical example} \label{Example}

\begin{figure}[h]
\begin{minipage}[b]{1\textwidth}
\subfigure[Cylindrical bar axially loaded
   \label{fig1a} ]{ \includegraphics[width=.2\textwidth]{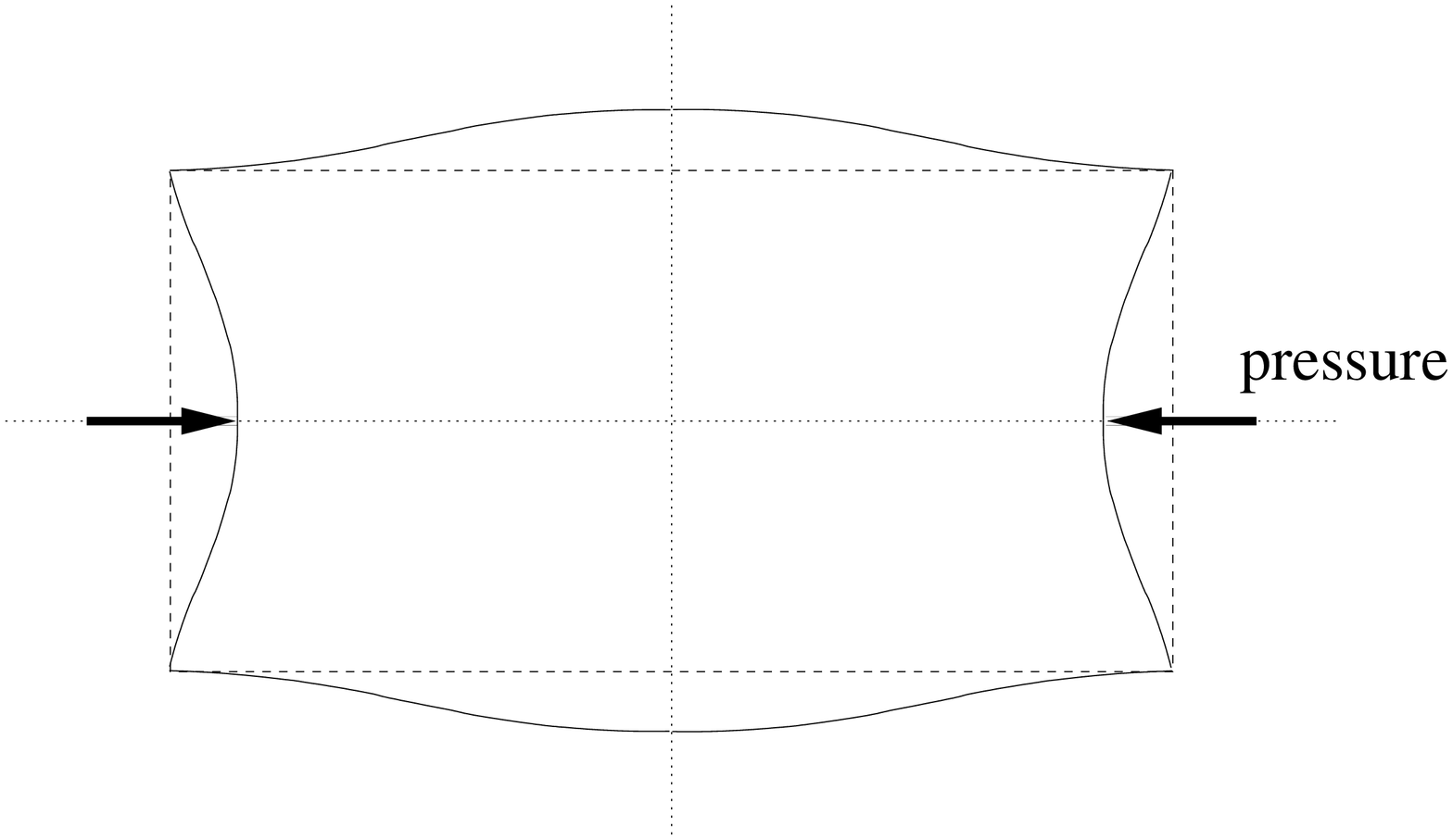}
   }
\subfigure[Oblique intersection
   \label{fig1b} ]{ \includegraphics[width=.2\textwidth]{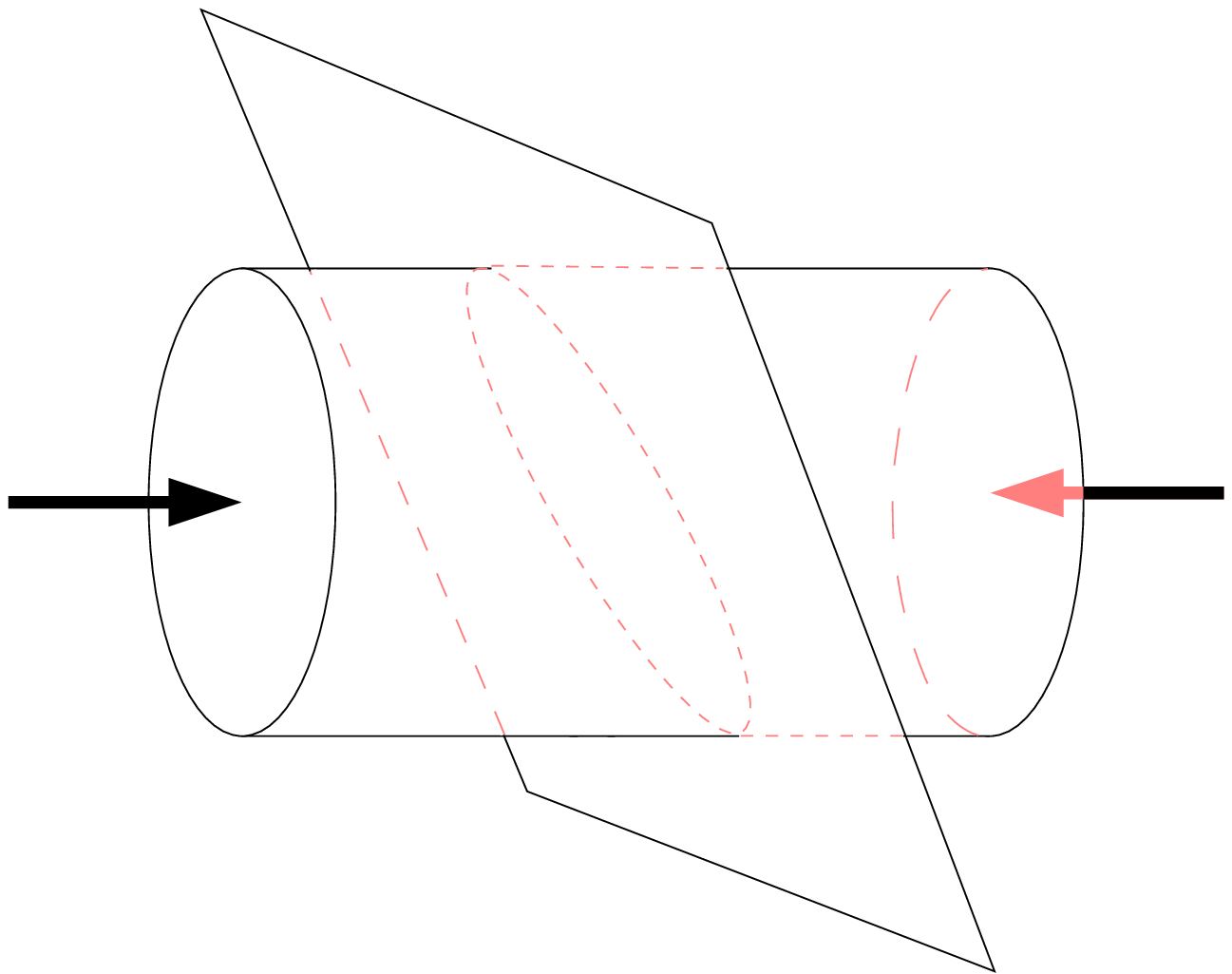} }
\end{minipage}
\begin{minipage}[b]{1\textwidth}
\subfigure[Original tensor component $A_{\bm{\eta} \bm{\eta}}$
   \label{fig1c} ]{ \includegraphics[width=.2\textwidth]{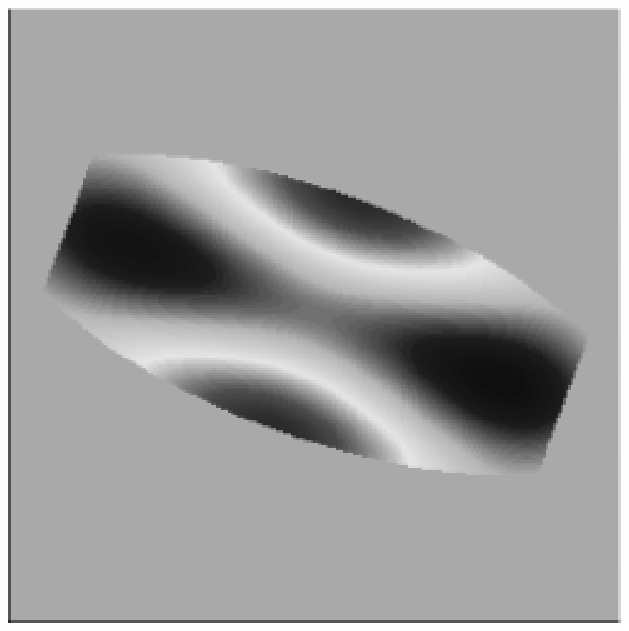}
   }
\subfigure[Reconstruction with $254\times 254$~pixels and $36$ scans
   \label{fig1d} ]{
   \includegraphics[width=.2\textwidth]{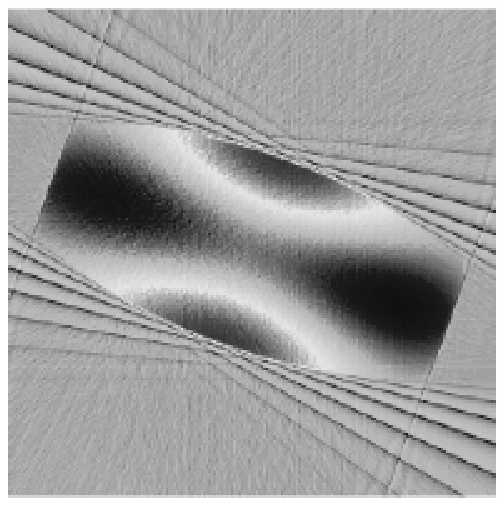} }
\end{minipage}

\caption{The reconstruction of the anisotropy tensor~(\ref{scale5a})
gives valuable information about the internal structure of the object,
even if the average dielectric constant $\epsilon$,
eq.~(\ref{average1}), is not known. \label{figure1}}
\end{figure}

Here we present a numerical example of reconstruction of a single tensor
component $A_{\bm{\eta} \bm{\eta}}$ for a plane intersecting the
object at an oblique angle. The polarization transformation data are
obtained from an artificial stress model of a cylindrical bar with a
circular cross-section which is subject to axial load and in turn
bulges out in the middle section, see Fig.~\ref{fig1a}.

Based on these artificial forward data we then employ our method and
reconstruct the ``normal'' component $A_{\eta\eta}$ on a plane passing
through the center of the cylinder and making an angle of $22^{\circ}$
with the symmetry axis, see Fig.~\ref{fig1b}. The original plot of
$A_{\eta\eta}$ in this plane is shown in Fig.~\ref{fig1c}; the
reconstructed image in Fig.~\ref{fig1d} has $254\times 254$ pixels,
assuming that $36$ scans, one scan on every $5^{\circ}$, have been
performed in the plane. It will be seen that the reconstruction
contains artefacts which are typical of a Radon inversion; they can be
reduced by increasing the number of scans, e.g., by performing one
scan on each degree, for a total of $180$ scans. The result in this
case is almost indistinguishable from the original in Fig.~\ref{fig1c}
so that we have refrained from showing it.

For the inverse Radon transformation, the Matlab function {\sf iradon}
has been used which utilizes a filtered back-projection
algorithm\cite{Natterer}.

\section{Summary}

We have presented a novel way to reconstruct an arbitrarily
inhomogeneous anisotropic dielectric tensor inside a transparent
non-absorbing medium, under the conditions that  the birefringence is
a deviation from a homogeneous isotropic average, and that this
deviation is weak. It was shown that the associated linearized inverse
problem of reconstructing the dielectric tensor from polarization
transformation data gathered by optical-tomographical means can be
reduced to six scalar Radon inversions which allow the determination
of the permittivity tensor completely. We also supplied a more
rigorous derivation of the usual equations of integrated
photoelasticity which define the inverse problem for the dielectric
tensor; our exposition was based on a careful description of the
various approximations that enter the derivation of the
photoelasticity equations from Maxwell's equations in a material
medium.

\section*{Acknowledgements}

The authors acknowledge support from EPSRC grant~GR/86300/01.

\end{document}